\documentstyle[art8,aaspp4]{article}

\def\ltsim{ \,{}^<_\sim\, }
\def\gtsim{ \,{}^>_\sim\, }
\def\etal{et~al.}

\lefthead{Harris \etal }
\righthead{NGC 4874 Globular Clusters}

\begin{document}
\large

\title{The Globular Cluster Systems in the Coma Ellipticals.  II:
Metallicity Distribution and Radial Structure in NGC 4874, and Implications
for Galaxy Formation\footnote{Based on observations 
with the NASA/ESA {\it
Hubble Space Telescope}, obtained at the Space Telescope Science
Institute, which is operated by the Association of Universities for
Research in Astronomy, Inc., under NASA contract NAS 5-26555.} }

\author{William E.~Harris and J.~J.~Kavelaars}
\affil{Department of Physics and Astronomy, McMaster University, 
Hamilton ON L8S 4M1, CANADA; harris,kavelaar@physics.mcmaster.ca}

\author{David A. Hanes}
\affil{Department of Physics, Queen's University, Kingston ON K7L 3N6, CANADA; hanes@astro.queensu.ca}

\author{James E. Hesser}
\affil{Dominion Astrophysical Observatory, Herzberg Institute of Astrophysics, National Research Council, 5071 West Saanich Road, Victoria BC V8W 3P6, CANADA; james.hesser@hia.nrc.ca}

\and

\author{Christopher J. Pritchet}
\affil{Department of Physics and Astronomy, University of Victoria, Box 3055, Victoria, BC V8W 3P6, CANADA; pritchet@phys.uvic.ca}

\begin{abstract}
Deep HST/WFPC2 images in $V$ and $I$ are used to
investigate the globular cluster system (GCS)
in NGC 4874, the central cD galaxy of the Coma cluster.  
Although the luminosity function of the clusters displays its normal
Gaussian-like shape and turnover level, other features
of the system are surprising.   We find the GCS to be (a) spatially 
extended, with core radius $r_c \sim 22$ kpc, (b) entirely
metal-poor (a narrow, unimodal metallicity distribution with
$\langle$[Fe/H]$\rangle \sim -1.5$),
and (c) modestly populated for a cD-type galaxy, with
specific frequency $S_N = 3.7 \pm 0.5$.
Model interpretations suggest to us that as much as half of this
galaxy might have accreted from low-mass satellites, but no single
one of the three classic modes of galaxy formation
(accretion, disk mergers, {\it in situ} formation) can supply a
fully satisfactory model for the formation of NGC 4874.  Even when
they are used in combination, strong challenges to these models
remain.  We suggest that the principal anomaly in this GCS is essentially the
complete lack of metal-rich clusters.  If these were present in
normal (M87-like) numbers in addition to the metal-poor ones
that are already there, then the GCS in total would more closely
resemble what we see in many other giant E galaxies.
This supergiant galaxy appears to have avoided forming
globular clusters during the main metal-rich stage of star formation
which built the bulk of the galaxy.  
\end{abstract}


\section{Introduction}

NGC 4874 is the central cD-type galaxy in the rich
Coma cluster.  As such, it 
presents a rare opportunity for us to study the characteristics
of a globular cluster system (GCS) in an extreme environment:
a host galaxy which is
near the very top end of the luminosity scale 
and is within an extremely rich environment of other galaxies.
The characteristics of the GCS including the metallicity distribution
(MDF) of the clusters, the specific frequency of the cluster system,
its spatial structure, and the luminosity distribution of the clusters,
all provide distinctive clues to the evolutionary history of the galaxy
(see, e.g., \cite{har91}, 1999 and \cite{ash98}
for extensive reviews).

In Paper I (\cite{kav99}), we described new imaging observations
of NGC 4874 taken with the WFPC2 cameras on board HST, and discussed the
luminosity distribution function (GCLF) of the globular clusters.
Here, we discuss the spatial distribution of the clusters, the MDF,
and the specific frequency, and we describe how they may constrain competing
scenarios for the formation history of this unusual system.

The raw WFPC2 data for this program (GO-5905) consisted of 
$V$ (F606W) exposures totalling 20940 sec, and $I$ (F814W)
exposures totalling 8720 sec. 
NGC 4874 was approximately centered in the PC1 chip to maximize the total
globular cluster population falling within our field of view.
A complete description of the data reduction procedures, including DAOPHOT
and ALLFRAME measurement, photometric transformations, and
artificial-star tests of the photometric error and completeness
functions, is given in Paper I.

\begin{figure}
\epsscale{0.8}
\plotone{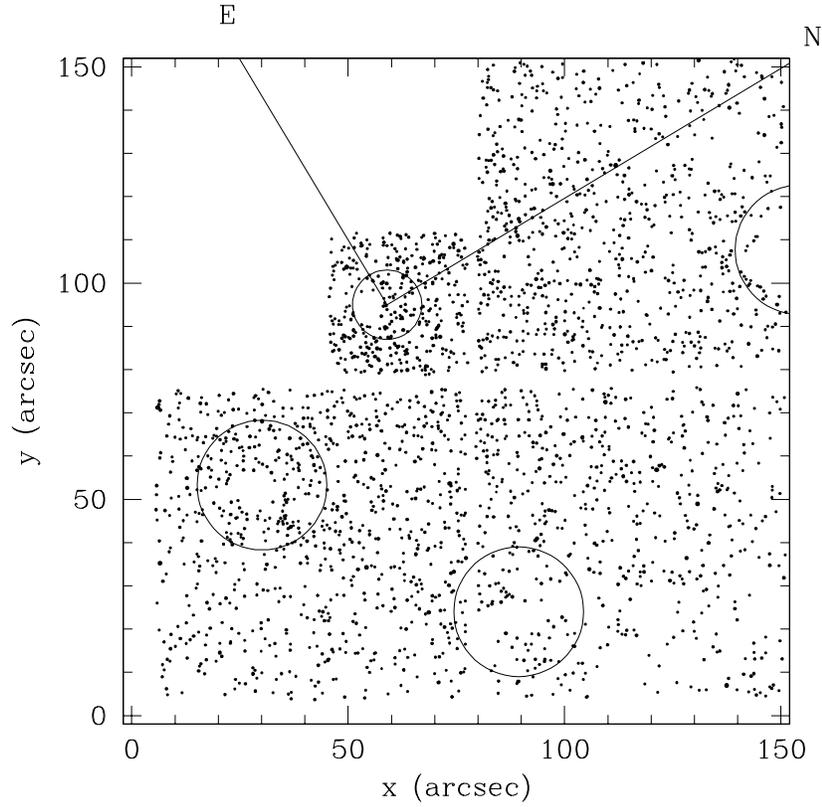}
\caption{Plot of the brighter starlike objects ($V < 27.0$) 
detected and measured in the NGC 4874 WFPC2 field.  The galaxy
center is at $(59'', 95'')$ near the center of the PC1 chip.  
Cardinal directions (north, east) relative to the center of
NGC 4874 are marked with the straight lines.  Objects regarded as
``nonstellar'' according to image structure analysis (see Paper I)
have been culled from the sample.  The area nearest the center of NGC 4874,
and three neighboring E galaxies marked with the circles (NGC 4871,
4872, 4873) were
masked out in the analysis, as were narrow areas of slight
vignetting along the four chip edges.}
\label{fig:xyplot}
\end{figure}

\placefigure{fig:xyplot}

In Figure \ref{fig:xyplot} 
we show the WFPC2 field with the locations of the
brightest detected objects. 
The $y-$axis of the array is directed $59\fdg0$ east of north.
In total, 4313 objects classified as
``starlike'' in structure (consisting of globular clusters around
NGC 4874, some faint very compact background galaxies, and a very few
foreground field stars) were measured in $V$. Of these, 3344 were
also measured in $I$ and thus $(V-I)$.
The concentration of objects around NGC 4874 is evident, but the
distribution is much less centrally concentrated than (for example)
around another large Coma elliptical, IC 4051 (\cite{woo99}).

In Paper I, the distribution in $V$ magnitude (the GCLF)
was employed to determine the distance modulus to the Coma cluster and
an estimate of the Hubble constant.  We found the GCLF ``turnover'' point
to lie at $V = 27.88 \pm 0.12$, leading to $\mu_0$(Coma) $= 35.05 \pm 0.12$
and $H_0 = 69 \pm 9$ km s$^{-1}$ Mpc$^{-1}$.  In the discussion of
this paper, we will simply assume a Coma distance $d = 100$ Mpc to convert
any angular measurements into linear ones.

\begin{figure}
\epsscale{0.8}
\plotone{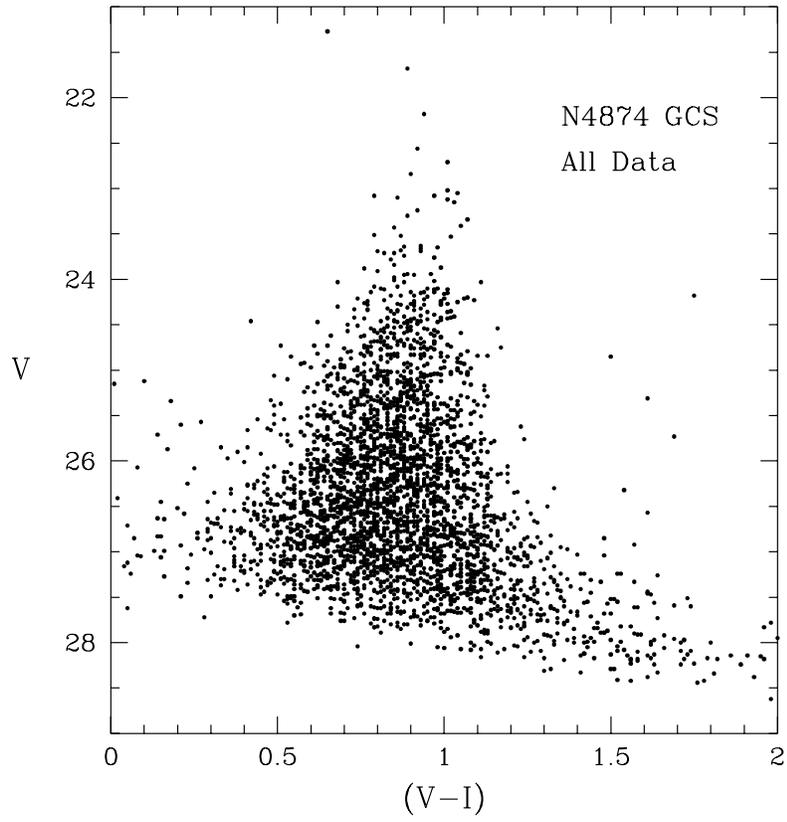}
\caption{Color-magnitude diagram for all starlike objects in the
NGC 4874 field with measured $(V-I)$ color indices.
The $I-$band exposures were significantly shorter than $V$, so
that the data for $I \gtsim 25$ are very uncertain.}
\label{fig:cmdplot}
\end{figure}

The color-magnitude diagram for all objects with measured $(V-I)$ color
indices is shown in Figure \ref{fig:cmdplot}. 
Since the $I$ frames had relatively short
total exposure time, this color distribution does not quite reach
the turnover point, and fainter than $V \sim 26$ the appearance of
the diagram is dominated by the random measurement uncertainty of
the photometry.  We will discuss the color (metallicity) distribution
in more detail in section 4 below.

\placefigure{fig:cmdplot}

\section{The Radial Distribution: A Galactic or Intergalactic GCS?}

\begin{figure}
\epsscale{0.8}
\plotone{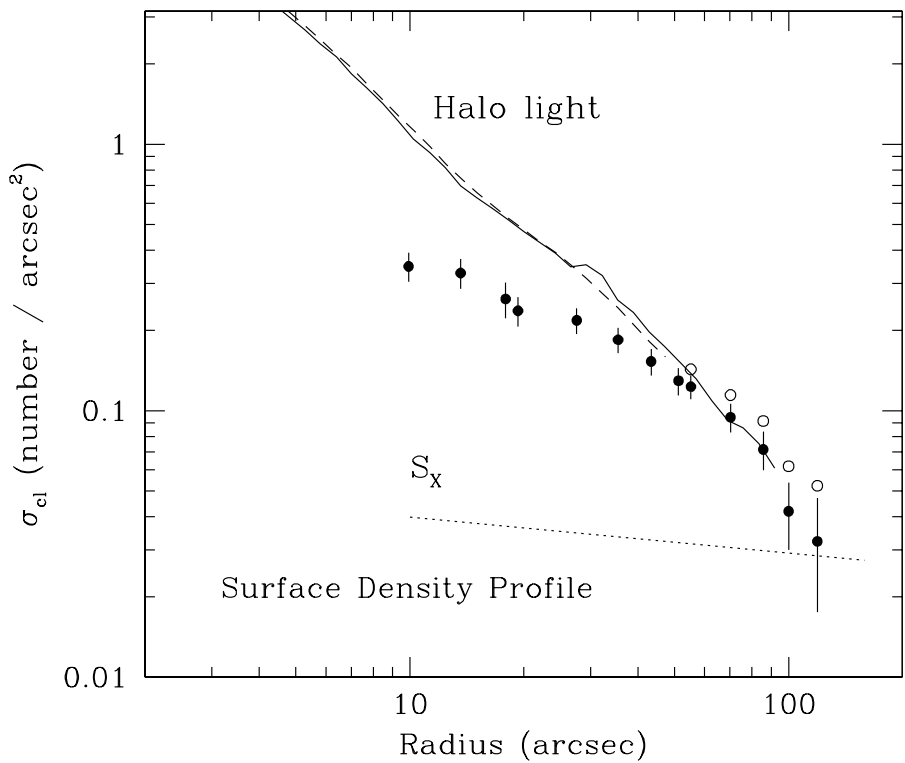}
\caption{Radial profile for various components of the NGC 4874 system.
{\it Solid dots} with error bars: the profile for the globular cluster
system (from Table 1, with $\sigma_{cl} = \sigma - \sigma_b$ and an
adopted background level $\sigma_b = 0.02$ arcsec$^{-2}$).
{\it Open circles} show the outermost five annuli as they would be
if the background density was arbitrarily assumed to be zero.
{\it Solid line:} halo light profile in the $R-$band,
from Peletier \etal\ (1990).  {\it Dashed line:}  halo light profile 
in the Gunn $r-$band, from Jorgensen \etal\ (1992).
{\it Dotted line:} surface intensity $S_X$ of the intracluster X-ray halo
gas in the Coma cluster (see text).  The light and X-ray profiles have
been arbitrarily shifted vertically for display purposes; see the
next figure for a comparison with proper normalization.}
\label{fig:profile}
\end{figure}

Pioneering attempts to measure the NGC 4874 GCS from ground-based
CFHT imaging (\cite{har87}; \cite{tho87}) succeeded in resolving the brightest
clusters and hinted that their radial distribution was
surprisingly flat, with the surface density of clusters $\sigma_{cl}$
falling off with radius as $\sim r^{-1}$ for $r \gtsim 20''$.
Our new HST data cover a similar radial range but of course
penetrate considerably deeper.
To obtain the radial profile of the GCS, we subdivided the WFPC2 field
into several annuli and calculated the number density of starlike
objects down to an adopted cutoff magnitude $V = 27$, above which 
the detection incompleteness was small.
The results are summarized in
Table \ref{tab:radprof} and shown in Figure \ref{fig:profile}.  
In the Table, successive columns give
(1) the mean radius of each annulus, (2) the number of starlike objects
within the annulus, after (small) completeness corrections, 
(3) the area $A$ of the annulus that falls within the WFPC2 boundaries,
and (4) the projected number density of starlike objects, $\sigma = n/A$.
Column (5) gives the deduced mass density after deprojection into
three dimensions, as described below.

The projected density $\sigma$ 
declines smoothly and slowly outward from the galaxy center
and is still declining
at the edges of our WFPC2 field ($r \sim 120''$ corresponds to almost 60
kpc linear radius).  Plainly, the GCS of this
giant galaxy spills past the borders of our field, making it difficult
to estimate the surrounding background level $\sigma_b$ 
from our data alone.  However, without knowing $\sigma_b$ we cannot
obtain the true profile of the GCS.  Lacking any true
``control'' field adjacent to NGC 4874, we have instead 
used data from another Coma field, the elliptical galaxy IC 4051
(\cite{woo99}), in which the magnitude limits,
data reduction procedures, etc., are all closely comparable to ours.
Fortunately, the IC 4051 GCS is much more centrally concentrated,
so that the starcounts in the
outer regions of its WFPC2 field are very close to the true surrounding
background.  We adopt $\sigma_b = (0.02 \pm 0.005)$ arcsec$^{-2}$ -- about
one-third the projected density of the outermost parts of our NGC 4874
field -- and subtract this from the entries in Table 1 to obtain the
profile plotted in Fig.~\ref{fig:profile}.

Notably, the slope of the $\sigma_{cl}(r)$ curve steepens outward,
continuously from $r \sim 10''$ (or $\simeq 5$ kpc) all the way
out to $r \sim 120''$ (60 kpc).  The flat inner-halo  core is 
remarkably extended, and the outer halo does not 
follow a single power-law exponent as well as in several
other giant ellipticals at the centers of rich clusters (e.g., Harris 1986,
1991; \cite{bla99}).  The steepening of the profile past
$r \sim 120''$ hints that the outer halo there may become truncated
(because of tidal interactions with the other large
Coma ellipticals?).  This dropoff in $\sigma_{cl}$ 
is not just an artifact of our adopted background level:  it
would still be present even if we had arbitrarily adopted $\sigma_b = 0$,
as shown by the open circles in Fig.~\ref{fig:profile}.

The profile of the halo {\it light} of NGC 4874 has
been measured via CCD photometry
to comparable radii by \cite{pel90}
and \cite{jor92}.  In Fig.~\ref{fig:profile}  we compare their data
with the GCS profile.
The halo surface intensity is well described by a simple power
law $\mu$(halo) $\sim r^{-1.1}$ over the range $5''\ltsim r \ltsim
90''$ (see particularly the Peletier \etal\ $R-$band data, which
cover the largest radial range.  NGC 4874 is well known to be a
difficult target for wide-field surface photometry, since at radii
beyond $100''$ or so, its halo light becomes confused with
the light from several neighboring large ellipticals in the Coma core
region.).  However, the GCS profile does not fit such a simple
description:  its slope steepens continuously outward, and it can
be claimed to match the halo light profile only for $r \gtsim 40''$.

Very few other giant ellipticals, even cD types, have GCS profiles
this extended.  Other roughly comparable cases may include
NGC 3311, the very diffuse central cD in 
Hydra I (\cite{har86}; \cite{mcl95}), perhaps NGC 6166, the cD in
Abell 2199 (\cite{pri90}), and some of the Abell-cluster central
giants studied by Blakeslee (1999) \nocite{bla99}, particularly
A754.  We have used standard \cite{kin66} model fits to the GCS radial
distribution in order to obtain formal estimates of the
{\it core radius} of the projected
profile.  Over a wide range of assumed central concentration
parameters, we find $r_c \simeq 45''$ (corresponding to 22 kpc).
By contrast, the core radius of the halo light is $\simeq
3\farcs3$ or 1.6 kpc (\cite{you79}).

\placetable{tab:radprof}

\placefigure{fig:profile}

With a halo cluster system this extended, it is reasonable to ask
whether or not the GCS belongs to the central galaxy, or instead to the
surrounding potential well of the Coma cluster as a whole. 
To address this question, we can use the profile of
the hot X-ray gas surrounding NGC
4874 as a reasonable tracer of the Coma potential
well.  Several sets of observations (e.g., \cite{hug89}; \cite{wat92};
\cite{whi93}; \cite{mak94}; \cite{dow95}) indicate that the
X-ray emission around the two central supergiants
NGC 4874 and NGC 4889 almost certainly arises from
intracluster material filling the Coma potential well. 
Within this central region there is no
smaller-scale gas component present that can clearly be 
identified with the individual galaxies.
Within our field of study ($r \ltsim 3'$), the gas is nearly isothermal
and its projected intensity is essentially constant, 
$S_X \sim r^{-0.13}$ (see, e.g.,
Figure 2 of Dow \& White). The core radius of the
X-ray distribution as a whole (\cite{hug89}; \cite{wat92}; \cite{mak94})
is $r_c \simeq 10'$, corresponding about 300 kpc.

The overall spatial distribution of the Coma galaxies also provides
a measure of the scale size of the potential well.  Quoting a single
typical scale size is difficult because of the well known
subclustering of the galaxies (especially the E and brighter dE
types:  e.g., \cite{biv96}, \cite{col96}), but representative estimates
of core radii are in the range $r_c \sim 10'$, very much like the
X-ray gas (\cite{ken82}; 
\cite{sec97}).  Secker \etal\ (1997) also find that the
scale radius for the {\it faintest} dE galaxies ($M_V \gtsim -15.4$)
is larger still, at $r_c \simeq 22'$.  Several authors have taken
this extended distribution, as well as the rather flat luminosity
function of the dE's, to suggest that these small cluster members
have been depleted by tidal disruption or accretion onto the giants
(e.g., \cite{ber95}; \cite{tho93}; \cite{sec97}; \cite{lob97}).
In summary, it appears that the scale radius of 
the Coma cluster potential well 
is about one order of magnitude larger than that of the GCS
around NGC 4874.  

These comparisons suggest to us that the
GCS should not be associated with the Coma cluster as a whole.  
Evidence more strongly favoring its identification
with the central galaxy is that the GCS profile  
(Fig.~\ref{fig:profile}) is similar in its outer regions to the
halo light profile of NGC 4874, continuing to steepen past the outer limits
of our data rather than flattening off as it would if an
intragalactic component made up a large part of the GCS.
Although we cannot rule out the existence of a more extended,
free-floating GCS component belonging to the broader Coma potential
well, without larger-scale data in hand it is not possible to
make any further tests for it.  In the Discussion below, we will return
to the issue of building NGC 4874 as a whole from accreted
material.

\section{Specific Frequency and Mass Ratios}

We can now use the radial profile information to estimate the
specific frequency of the GCS,
$$S_N \equiv N_t \cdot 10^{0.4(M_V^t + 15)} \, ,$$
where $N_t$ is the total population of globular clusters and
$M_V^t$ is the $V-$band integrated  magnitude of the galaxy.
This quantity, which has sometimes been taken as
an indicator of the global efficiency
of globular cluster formation over the lifetime of the parent
galaxy, differs among giant ellipticals by more than an order
of magnitude (\cite{har91}; \cite{har98}).  For the cD-type
galaxies in particular, $S_N$ is known to increase
systematically with such external characteristics as total
galaxy luminosity, the size of the surrounding cluster of galaxies,
and the X-ray halo gas mass (\cite{bla97}, 1999; \cite{har98}; \cite{mcl99}).

Interpreting the trends in specific frequency has for many years
been a key point in the debate over the evolutionary histories
of elliptical galaxies (see, for example, \cite{har81}, 1995, 1999;
\cite{vdb82} and several later papers; 
\cite{sch87}; \cite{ash92}; \cite{wes95};
\cite{ws95}; \cite{car98}; \cite{for97};
\cite{kis98}; \cite{cot98}; \cite{zep99}; \cite{bla99}, among others).
These papers variously propose that the
often-high specific frequencies in cD galaxies may have arisen from
higher than average globular cluster formation efficiency in shocked
merging gas clouds; or by accretion from neighboring galaxies; or by the
presence of intergalactic globular clusters in large numbers; or by
the loss of gas in high proportions after an initial protogalactic
cluster formation period.  Counterarguments exist to all options,
and no completely satisfactory path has yet emerged.

Early ground-based observations of the very brightest clusters
around NGC 4874 gave uncertain results for $S_N$:  
Harris (1987) found $S_N \gtsim 10$ similar to M87, while
Thompson \& Valdes (1987) suggested a much lower value,
$S_N \gtsim 4$.  Intermediate
values near $S_N \sim 7$ were later found by \cite{bla95},
also from ground-based imaging,
based on a combination of surface brightness fluctuation analysis
and direct resolution of the brightest globulars.

From our WFPC2 data, we calculate $S_N$ in two ways:

\medskip

\noindent (1) We first derive a ``semi-global'' specific frequency covering
only the area of our WFPC2 starcounts, which extend
from $r_{min} = 7\farcs84$ to 
$r_{max} \simeq 130''$.  From Table \ref{tab:radprof}, 
we multiply the observed surface
density $\sigma_{cl} = \sigma - \sigma_b$ (where $\sigma_b = 0.02 \pm
0.005$) by the complete area $\pi (r_{outer}^2 - r_{inner}^2)$ of each
annulus.  The residual total of $(1785 \pm 265)$ is then the number 
of globular clusters brighter than $V = 27.0$ over that radial range.
For a Gaussian luminosity function with turnover at $V^0 = 27.88 \pm 0.12$
and dispersion $\sigma_V = 1.49$ (Paper I), the directly observed number
must be multiplied by $(3.60 \pm 0.35)$ to obtain the total cluster
population over all magnitudes, yielding $N_t = 6425 \pm 1150$.

Over the {\it same} radial region, we also integrate the $R-$band
light profile of Peletier \etal\ to obtain $R$(total) = 11.26,
or $V$(total) $= 11.86 \pm 0.05$ for a typical gE color index
$(V-R) = 0.60 \pm 0.05$ (e.g., \cite{but95}; \cite{pru98}).
With $(m-M)_0$(Coma) $=35.03 \pm
0.12$ (Paper I), we obtain $M_V^t = -23.17 \pm 0.13$ for the total
$V$ luminosity of the enclosed light.  The resulting specific frequency
is $S_N = 3.5 \pm 0.7$, where the quoted error margin includes the
statistical uncertainty in the number of clusters as well as the uncertainties
in the turnover luminosity and the distance modulus.

\medskip

\noindent (2) Next we calculate a global specific frequency by
estimating the total cluster population over all radii and dividing
by the integrated luminosity of the entire galaxy.  For the innermost region
$r < 7\farcs84$, recognizing that the GCS profile is nearly flat we
assume $\sigma_{cl} \simeq 0.4$ arcsec$^{-2}$ there, which translates
into $\sim 75 - 80$ clusters brighter than $V=27.0$.  This is
only a 5\% addition
to the directly observed sum given above.  The outward extrapolation
of the GCS is, however, less certain:  at radii beyond $r
\simeq 130''$, we do not know how steeply the GCS profile continues to
fall off.  Using the last three observed points from Figure \ref{fig:profile}
as a guide, we will assume $\sigma_{cl} \sim r^{-2}$ 
and (more or less arbitrarily)
truncate it at $r = 400''$ ($\sim 200$ kpc) where we are fully into
the larger-scale Coma galaxy environment.  
This procedure yields a further,
very uncertain, $\sim 700$ clusters.  The total over all magnitudes and
all radii is then $N_t \simeq 9200 \pm 1500$. 

The integrated magnitude of NGC 4874 is $V^t = 11.68$ (RC3 catalog value)
or $M_V^t = -23.39 \pm 0.13$.  We then obtain a global specific frequency
$S_N = 4.1 \pm 0.7$.  This total will, of
course, be an overestimate {\it if} the outer radial profile continues to
steepen past our last observed point.  Tidal truncation of the NGC
4874 halo is likely to be imposed at a radius somewhere near
$r \sim 3'-4'$ by the similarly large supergiant
NGC 4889, which is only $7\farcm3$ (220 kpc) away projected on the
sky.

Three neighboring large S0 galaxies (NGC 4871, 4872, 4873) also
appear in our WFPC2 field of view.
No obvious globular cluster populations
were visible around any of these, but as a precaution, 
circles of radius $15''$ around each one were masked out of our data 
(see Fig.~\ref{fig:xyplot}).  All 
three of these galaxies have
integrated magnitudes $V_T \simeq 14.2$ ($M_V \sim -21$, one
order of magnitude less luminous than NGC 4874).  If
they have roughly normal specific frequencies $S_N \simeq 4$, we
would expect each to contribute $\ltsim 50$ globular clusters
falling outside the masked-out circles and
brighter than our photometric limit of $V=27$. 
Thus these would
contribute {\it at most} 9\% of our measured residual total $\sim
1800$ clusters in the entire field. 
No adjustments to our measured $S_N$ have been applied for this effect.

\medskip

The two approaches are basically in good agreement, and for the
following discussion we will adopt $S_N = 3.7 \pm 0.5$.  It is quite plain
from either set of assumptions that {\it NGC 4874 is not a ``high
specific frequency'' giant} like M87 and some other cDs.
Instead, $S_N \sim 4$ is entirely similar to the
run-of-the-mill (non-central) ellipticals in clusters such as Virgo and
Fornax (\cite{har91}, 1999) and places it at an anomalously low
position relative to other cD's in rich clusters.  For example, the
correlation of $S_N$ with parameters such as the luminosity of the
parent cD, or the velocity dispersion of the surrounding galaxy
cluster (\cite{bla97}; \cite{har98}) would lead us to expect a global 
specific frequency for NGC 4874 about twice as large as it is.

We might reasonably ask why the estimates from the
previous ground-based observations
(see above) were different.  Neither of the CFHT results (\cite{har87},
\cite{tho87}) presents much cause for serious concern, since 
the relatively bright limiting magnitudes in those early studies led to
uncomfortably large extrapolations
to predict the total cluster population over all magnitudes,
The estimated specific frequencies were thus 
extremely sensitive to both the assumed background levels and the 
assumed characteristics of the Gaussian GCLF.  
Under these circumstances, factor-of-two discrepancies can easily arise.

The \cite{bla95} result
deserves a closer comparison, since it
is based on a carefully executed combination of SBF signal 
measurement and direct resolution of the brightest clusters. 
Their weighted average specific frequency over four annuli covering
the radial range $11'' < r < 175''$ (see their Table 2) 
is $S_N \simeq 6.3 \pm 0.5$.
Matching their analysis with ours reveals two points of note:
(a) Blakeslee \etal\ adopted a Coma distance modulus 
$\simeq 0.35$ mag smaller than ours, and a slightly smaller value
for the Gaussian dispersion of the GCLF (1.4 vs.~1.49).  Adjusting
these parameters back to our adopted values turns out to lower their
estimated $S_N$ by about 25\%, bringing it well within the 
uncertainty range of our measurement.
(b) Their number of {\it directly resolved} bright clusters is, in fact,
similar to ours if normalized to the same magnitude range.
Within their observed radial range, they find 
a residual total of $\sim 140$ clusters brighter than $I = 23.7$
over the radial range $0\farcm22 - 2\farcm92$
(after correction of each annulus to the same magnitude limit).
By comparison,  
in our data there are roughly 140 objects brighter than this
limit over very nearly the same radial range, $0\farcm13 -
2\farcm17$.  In summary, there appears to be no basic 
disagreement between their study and ours at the $\sim25$\% level
mentioned above.

\begin{figure}
\epsscale{0.8}
\plotone{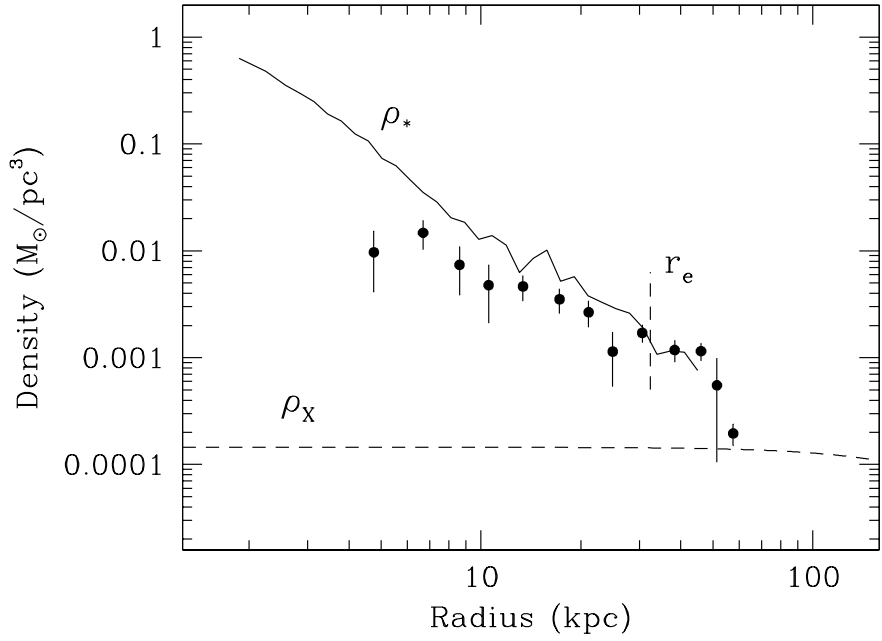}
\caption{Volume density profiles for three components of the NGC
4874:  the halo light (solid line, labelled $\rho_{\star}$), the
globular cluster system (solid dots), and the X-ray halo gas
(dashed line, labelled $\rho_X$).  The globular cluster density
profile has been divided by a mass ratio
$\epsilon = M_{GCS}/M_g = 0.003$ to normalize it to the halo light;
see text for discussion.  The effective radius of the galaxy
light, $r_e \simeq 66'' \simeq 32$ kpc (Peletier \etal 1990) is marked.}
\label{fig:3Dprofile}
\end{figure}

Another way to describe the total cluster population which is
more directly relevant to the physical formation efficiency of
globular clusters is the {\it mass ratio} in the GCS relative to
the other halo components (e.g., \cite{har98}; \cite{mcl99};
\cite{jj99}).
In Figure \ref{fig:3Dprofile}, we explicitly compare the mean mass
density profile in the GCS with those of 
he halo field stars and the hot X-ray gas.

To obtain the three-dimensional density profile of the halo light,
we deprojected the halo surface brightness in $\mu_B$ (Peletier \etal\ 1990)
in the manner described by McLaughlin (1999) and 
converted it to mass per unit volume with the assumption $(M/L)_B = 8$
for an old-halo stellar population (e.g., \cite{vdm91}).
The GCS profile was also deprojected with McLaughlin's algorithm, corrected
for clusters fainter than our photometric limit $V=27$ (see below),
and multiplied by an assumed mean cluster mass $\langle M_{cl} \rangle
= 2.5 \times 10^5
M_{\odot}$.  In both cases we assumed spherical symmetry to perform
the deprojection, along with a surface density profile $\sigma
\sim r^{-2}$ at radii beyond our outermost measured annulus (though
the shape of the deprojected density profile is not at all sensitive to this
latter assumption except for the outermost observed point).
Lastly, the X-ray density profile is from Makino (1994,
adjusted to a distance scale $H_0 = 70$).  Because of its huge radial
extent, the gaseous component is at nearly constant 
density over our region of observation.

To normalize $\rho_{GCS}$ properly to the other components, we divide
it by a ratio $\epsilon$ which is the
ratio of total mass in the GCS to the total mass in the galaxy
within the same radial region, $\epsilon = M_{GCS}/M_g$.
(The ``efficiency'' ratio $\eta_{GC}$ used by \cite{bla97} and
\cite{bla99} can be converted to $\epsilon$ by multiplying it by
the mean globular cluster mass $M_{cl}$.  We prefer to use 
$\epsilon$, since it is a strictly dimensionless quantity which can easily
be interpreted as the fraction of star-forming mass converted to
clusters.)  Here,
the galaxy mass $M_g$ is to be thought of as comprising both the
mass in the halo stars and in the X-ray gas, $M_g = M_{\star} + M_X$,
and it implicitly assumes that the gas mass within the region
represents either protogalactic material that was unused for 
star formation, or processed material that was
later ejected through supernovae,  stellar winds, or tidal
stripping.  (See the discussions of McLaughlin 1999,  Harris \etal\ 1998,
or \cite{bla99}.
In the case of NGC 4874, this assumption may not be correct since
the gas is associated with the Coma potential rather than one
specific galaxy.  However, the distinction is unimportant since
$M_X$ turns out to be small; see below.)

In essence, $\epsilon$ is then an estimate of the global mass formation
efficiency ratio for the globular clusters.
To avoid a possible bias in $\epsilon$ from the fact that 
globular clusters in the inner core of the galaxy will have been
preferentially destroyed by dynamical effects (dynamical friction,
tidal shocking, and tidally enhanced evaporation), McLaughlin
defines this mass ratio for $r \gtsim r_e$ (the de Vaucouleurs
effective radius), outside of which these
dynamical effects are small.  For NGC 4874, the effective radius
is, however, quite large ($r_e = 66''$; Peletier \etal\ 1990). 
This restricts us to only the outermost five annuli in our GCS profile and
unfortunately leaves only a small overlap with the halo light
profile.  Nevertheless, under these conditions we find that we
need to adopt $\epsilon \simeq 0.003 \pm 0.0005$
to bring $\rho_{GCS}$ into alignment with ($\rho_{\star} + \rho_X$).
In this radial region ($r \ltsim 55'' - 110''$), the halo
light contributes 90\% of $M_g$ and the X-ray gas only 10\%.

Our estimate of $\epsilon$ is similar to McLaughlin's ``universal'' 
$\langle \epsilon \rangle = 0.0026$ which he derived from the average
of M87, NGC 4472, and NGC 1399.  
Blakeslee (1999) \nocite{bla99} provides additional evidence 
that $\epsilon$ takes on a similar value in
several other brightest-cluster ellipticals.
As expected, at smaller radii we see that 
the GCS profile gradually falls below the halo light.  
Bearing in mind the 
uncertainties in the various conversion parameters (particularly the
adopted mass-to-light ratios for the clusters and the old-halo
light), we therefore suggest that   
NGC 4874 was roughly as effective at forming globular
clusters as were the Virgo and Fornax ellipticals, and other BCG's.

\placefigure{fig:3Dprofile}

\section{Color and Metallicity Distribution}

The final information we need to add to our discussion is the color
or metallicity distribution of the GCS.  Broad or bimodal MDFs
are found in giant E galaxies about half the time (e.g., 
\cite{kw99}) and are conventionally taken to signal a
complex or multi-phase formation history (e.g., \cite{ash92};
\cite{for97}; \cite{kis98}; \cite{glh99}).
In addition, complex MDFs are certainly more common among the more
luminous ellipticals.
Here again, however, NGC 4874 turns out to give us a surprise.

\begin{figure}
\epsscale{0.8}
\plotone{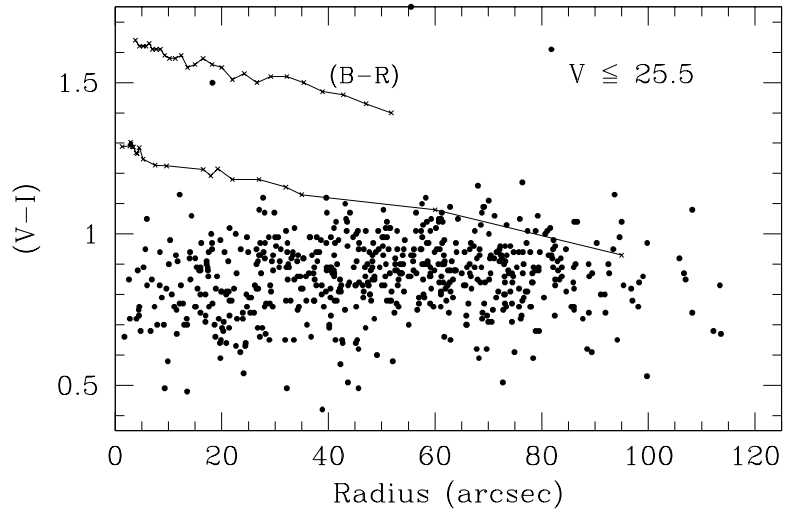}
\caption{Color index $(V-I)$ versus radius for the brightest
globular clusters in NGC 4874 ($V \leq 25.5$).
{\it Upper solid line:} Color profile in $(B-R)$ for the halo light
of NGC 4874, from Peletier \etal (1990).  {\it Lower solid line:}
Our measurement of the $(V-I)$ halo color; see text.}
\label{fig:radcolor}
\end{figure}

\placefigure{fig:radcolor}

Our information about the MDF relies heavily on the relatively
small number of clusters with small photometric errors, i.e., those
brighter than $V \sim 26$ 
(although there is no evidence that
mean color is correlated in any way with luminosity; 
see Fig.~\ref{fig:cmdplot}).
The distribution of $(V-I)$ with radius $r$ for the brightest
objects is displayed in Figure \ref{fig:radcolor}.  
Two results can immediately be drawn
from this plot:  (a) There is little if any gradient of mean
cluster color with radius.\footnote{Nominally, the clusters in the
centermost $ \simeq 25''$ have a slightly {\it bluer} mean color than those
at larger radii.  Although
this trend looks intriguing, we cannot ascribe any great
significance to it since that
radius corresponds to the transition zone between the PC1 chip and
the WF2,3,4 chips. 
As noted below, aperture corrections and photometric zeropoint
differences may produce chip-to-chip differences of
up to 0.05 mag that are hard to trace.  
Measurements of the cluster colors with a more
sensitive color index will be needed to investigate the
reality of this effect.}
(b) The range of colors within the GCS is {\it narrow} and
relatively {\it blue}.  All of these results are unexpected to
varying degrees.
For 146 objects with photometric uncertainties $\sigma(V-I) \leq
0.07$, the mean color is $\langle V-I \rangle = 0.907 \pm 0.008$,
with an rms dispersion $\sigma_{V-I} = 0.093$.  

\begin{figure}
\epsscale{0.8}
\plotone{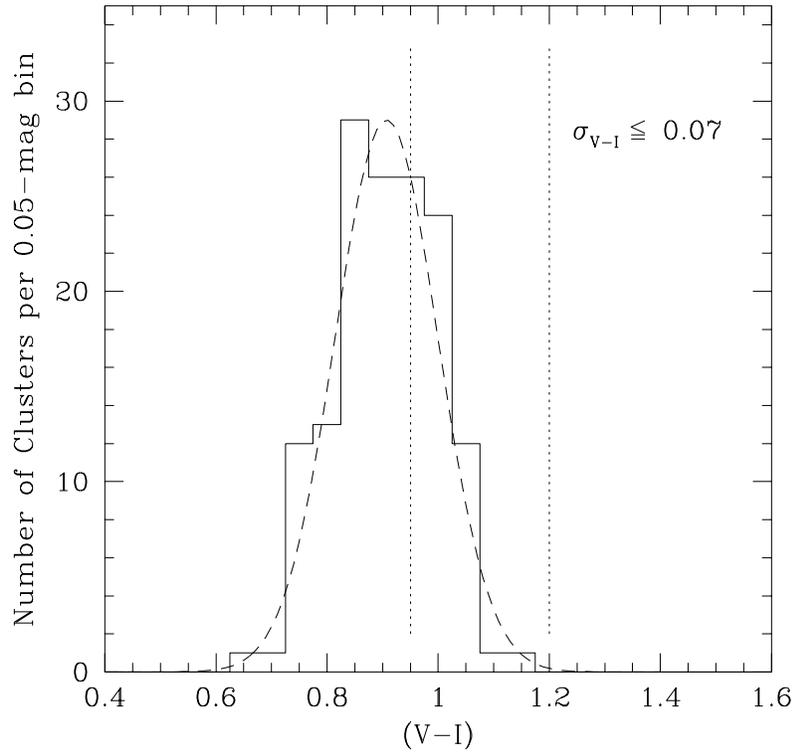}
\caption{Distribution in $(V-I)$ of the brightest clusters in NGC
4874.  A total of 146 objects with photometric uncertainty $\sigma_{V-I} \leq
0.07$ are shown, as a histogram of number per 0.05-mag bin.
{\it Dashed line:} Gaussian curve with mean $(V-I) = 0.907$ and
standard deviation $\sigma = 0.093$, the same as the raw histogram.
{\it Vertical dotted lines:} The lines at $(V-I) = 0.95$ and 1.20
represent the approximate locations of the two modes found in M87
and other Virgo and Fornax galaxies; see text.}
\label{fig:VIhisto}
\end{figure}

In Figure \ref{fig:VIhisto}, the 
color distribution is plotted in histogram form, along
with a single Gaussian curve with the same mean and standard
deviation as given above.  A simple Gaussian provides an adequate
match to the raw histogram, and clearly we cannot reject the hypothesis
that the MDF is unimodal.  
The average photometric measurement uncertainty is $\pm 0.064$
for this sample; subtracting it in quadrature from the observed
width $\sigma_{V-I} = 0.093$ then suggests that the 
intrinsic cluster-to-cluster scatter in color is at the level 
of $0.06 - 0.07$ mag.

\placefigure{fig:VIhisto}

\begin{figure}
\epsscale{0.8}
\plotone{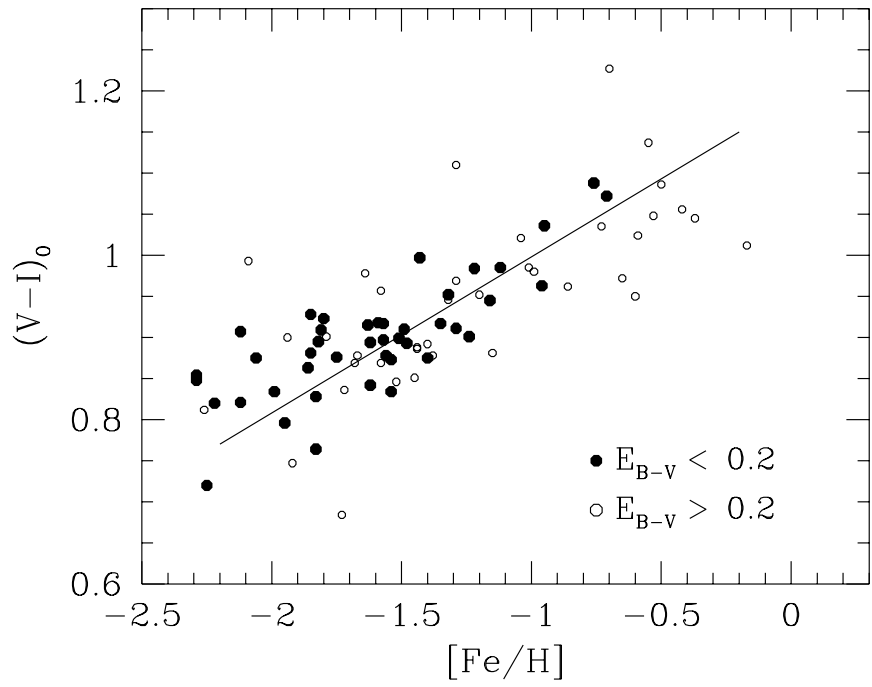}
\caption{Calibration of $(V-I)$ color index against metallicity
[Fe/H] for the Milky Way globular clusters.}
\label{fig:VIcalib}
\end{figure}

If we can safely regard the GCS as a conventional ``old halo'' 
component (i.e.,
having formed in the first few Gyr of the galaxy's history; see
below) then the photometric color can readily be translated into
metallicity.  A new calibration of [Fe/H] against $(V-I)_0$ from
the Milky Way globular clusters, with data drawn from the 1999
edition of the McMaster catalog (\cite{har96}, accessible at
{\tt http://physun.physics.mcmaster.ca/Globular.html}), is shown
in Figure \ref{fig:VIcalib}.  A conversion 
factor $E_{V-I} = 1.3 E_{B-V}$ is used to
deredden the integrated colors.  In Fig.~\ref{fig:VIcalib}, solid dots are 
the integrated colors of clusters
with foreground reddenings $E_{B-V} < 0.2$,
while open symbols are those with $0.2 < E_{B-V} < 0.6$.
Giving more weight to the lower-reddening clusters, we adopt a mean relation
$$ (V-I)_0 = 0.17 \, {\rm [Fe/H]} \, + \, 1.15 \, . $$
This relation is slightly steeper than the one derived by 
Kissler-Patig \etal\ (1997), namely $(V-I)_0 = 0.15$ [Fe/H] + 1.13, 
but shallower than an earlier calibration of \cite{cou90},
$(V-I)_0 = 0.198$ [Fe/H] + 1.207.  
Fortunately, the mean color of the 
NGC 4874 clusters is near the middle of the calibration line, where
all of the various conversion relations give similar results.

The sample mean $\langle V-I \rangle = 0.907$, with $E_{V-I}$(Coma)
= 0.01, gives $\langle$[Fe/H]$\rangle = -1.5 \pm 0.05$ (internal
uncertainty only).  If the external calibration uncertainty of the
$(V-I)$ scale is near $\pm 0.05$ (Paper I) then the true uncertainty
in this mean metallicity is near $\pm0.3$ dex.
Similarly, the intrinsic color dispersion of
$\sim 0.06 - 0.07$ corresponds to a metallicity histogram
width $\sigma$[Fe/H] $\simeq 0.4$.
Both the rather low mean metallicity and moderately narrow
dispersion strikingly resemble the analogous values for the halo
globular clusters in the Milky Way (e.g., \cite{zin85}; \cite{har99})
and in dwarf ellipticals (\cite{mil99}; \cite{har91}, 1999).
The mean [Fe/H] is also within the range typically occupied by
the metal-poor ``mode'' in giant ellipticals with bimodal
MDFs (e.g., \cite{for97}).  For comparison, we indicate in 
Fig.~\ref{fig:VIhisto} the $(V-I)$ colors 
for the metal-poor and metal-rich halves of
the MDFs in M87, NGC 4472, and in several Fornax ellipticals
(\cite{whi95}; \cite{kun99}; \cite{kis97}; \cite{puz99}).
To within the zero-point uncertainty of our $(V-I)$ color scale
(see the cautionary comments below), the lower-metallicity 
line is similar to the color of the NGC 4874 system.  
Any metal-rich component is entirely missing, even in the
core region.

\placefigure{fig:VIcalib}

To turn this very metal-poor MDF into a metal-rich one resembling
those found in M87 and other giant E galaxies (at [Fe/H] $\sim -0.5$),
we would have to claim that our $(V-I)$ scale zeropoint is wrong
by as much as $0.15 - 0.2$ magnitude, which we regard as extreme.
To verify our $(V-I)$ scale, we performed some 
additional consistency checks.  We took measurements of the surface
intensity of the NGC 4874 halo light at several locations
across the WFPC2 field.  The outer edges
of the field were used to define background, and the residual
light at each location was
translated through the photometric
calibration equations of Holtzman \etal\ (1995b; see
Paper I) to generate $(V-I)$ integrated colors.
The results are shown in Fig.~\ref{fig:radcolor} and listed in 
Table \ref{tab:surface}.
The typical uncertainty of each individual measure is $\pm 0.015$ mag.
Although no published measurements for the Coma
galaxies in $(V-I)$ are available
in the literature for comparison, it is encouraging that our
data give $(V-I) \simeq 1.2$ for the core of NGC 4874, 
which is exactly in the standard range for normal elliptical
galaxies (\cite{but95}). 

An additional result of interest
on its own merit is the color gradient of the halo, given roughly by
$ \Delta (V-I)/\Delta({\rm log} r'') = -0.17$.
From Figure 5, we see that the color of the outer halo
converges to the mean color of the GCS for $r \gtsim 100''$,
suggesting that they have comparably low metallicities there.
A gradient of similar nature was also found by \cite{pel90}
in the slightly more metallicity-sensitive index $(B-R)$
(also shown in Fig.~\ref{fig:radcolor}).

The surface photometry provides a consistency test of the accuracy
of the calibration equations, but the stellar photometry of the
individual objects involves the additional step of {\it aperture
corrections}.  Faint objects are best measured via PSF fitting or
through small apertures, but these must then be normalized to the
larger aperture of $0\farcs5$ radius that is used in
the \cite{hol95b} transformation equations.  The differences between
small and large apertures can be estimated through standard curves of
growth for the WFPC2 filters (e.g., \cite{hol95a}; \cite{suc97}).
These corrections may differ slightly -- typically by a few
hundredths of a magnitude -- from one CCD to the next 
on the WFPC2 array, from short to long exposures, and from one filter to
another.  

In our data reduction, we determined individual empirical PSFs 
whose instrumental magnitude scales were internally
equivalent to an aperture radius $r = 0\farcs2$ (or 2 pixels on the
WF2,3,4 fields).  We also independently derived aperture corrections
from the PSF scale to direct aperture-photometry magnitudes with
$0\farcs2$ radius, and
from $0\farcs2$ to $0\farcs5$, using isolated stars on each CCD. 
These adopted corrections are listed in Table 3.  
Aperture correction values from Holtzman \etal\ (1995a) and
Suchkov \& Casertano (1997), obtained from
much shorter-exposure standard-star fields, are given for comparison.
Although the $V$ offsets in all three studies are similar to within
$\pm 0.03$ mag, the discrepancy in $I$ between our work and
theirs is larger, by almost 0.1 mag.
However, the most important quantity for checking the 
accuracy of the $(V-I)$ color scale
is the {\it difference} $\Delta(V-I)$ between the aperture
corrections.  Here, our average for the WF chips
is $\Delta(V-I) = -0.06$, Holtzman \etal\ give $-0.04$,
and Suchkov \& Casertano give $+0.025$.  Inspection of the
relevant tables in their papers indicates that all these
values are typically uncertain by $\pm 0.03$ mag, so they are
not strongly in disagreement.  It is also worth noting that
if we had adopted the normalizations of either of these
other studies, our $(V-I)$ scale would have been even bluer.  

Nevertheless, it would clearly
be highly desirable to obtain additional photometry
of this system in a color index which is much more sensitive to
cluster metallicity than $(V-I)$ and thus more robust against
small zeropoint errors:  finer structure in
the MDF may be revealed,
and the metallicity peak can be independently checked.

\placetable{tab:surface}

\placetable{tab:apcorr}

\section{Discussion:  Alternatives for Galaxy Formation}

The globular cluster system in NGC 4874 presents us with an
unanticipated mix of characteristics:

\begin{itemize}
\item 
A ``normal'' GCLF (that is, a luminosity distribution with
Gaussian-like shape in number per unit magnitude, and a turnover
luminosity at the expected level for giant ellipticals);
\item
A spatial distribution with low central concentration and
extremely large core radius $r_c \simeq 22$ kpc;
\item
A specific frequency $S_N \simeq 4$, very close to the normal
value for giant ellipticals that are not cD's or brightest cluster
members;
\item
A narrow and surprisingly metal-poor metallicity distribution,
$\langle$[Fe/H]$\rangle \sim -1.5$ and $\sigma$[Fe/H] $\simeq 0.4$.
\end{itemize}

\noindent As indicated earlier, this combination is unexpected for
a cD-type galaxy in particular.  The primary goal of this work is to
help illuminate the sequence of events during galaxy formation, and
we now attempt to fit all of these characteristics into 
a consistent interpretive picture.

\subsection{Defining the Problem}

Competing ideas for the formation of E galaxies fall into three basic
categories:  traditional {\it in situ} formation from protogalactic gas;
growth by accretion or stripping 
of neighboring galaxies; and growth by merger of
gas-rich disk galaxies with accompanying star formation.

The data for NGC 4874 do not fit easily into any of these standard scenarios.
A basic merger-formation or accretion 
approach might initially seem most attractive for a cD-type galaxy:
recent numerical simulations for the way these galaxies build up
show that, because of
their privileged initial location near the center of a large
concentration of pregalactic material, they are likely to undergo 
a long series of accretions of ``fragments'' of many different
sizes, starting very early in the protogalactic epoch
and continuing for several Gyr 
(e.g., \cite{dub98}; \cite{wei96}; and references cited there).
Such a sequence would seem to provide the richest possible set
of opportunities for bringing in both metal-poor and metal-rich halo
material, supplies of gas for building new clusters, and growth of
the outer halo by harrassment and stripping of small neighbors
(\cite{moo96}).

However, the basic problem we encounter in all scenarios 
for the specific case of NGC 4874 is the lack of
metal-rich clusters in the present-day galaxy.
For example, the accretion model in the quantitative form
given by C\^ot\'e \etal\ (1998), or the accretion-by-harrassment
variation, assumes that an initial elliptical
gains material from smaller neighbors:  the original galaxy,
which is already moderately large,
generates the component of metal-rich clusters and stars during its
own single major formation burst, while the metal-poor ones
are added later from the smaller cannibalized satellites.  The 
puzzle we face is that in
NGC 4874 the metal-rich GCS component is entirely missing, even though
the bulk of the galaxy {\it light} is clearly red and metal-rich as in
normal ellipticals.  

In the merger picture (e.g., \cite{ash92} and subsequent papers),
the metal-poor clusters are assumed to belong to the progenitor disk
galaxies while the metal-richer ones are formed from the shocked gas
during the merger. (In the sense described by \cite{har99}, this process
can be thought of as an {\it active} merger, while pure accretion is a
{\it passive}, or gas-free merger in which new stars are not formed.)  
Since the starlight in the
resulting elliptical is predominantly metal-rich, most of its stars
should then have formed during the sequence of mergers.  Where, then, are
the metal-rich clusters that should have formed along with them?

Lastly, the {\it in situ} picture supposes that the broad or bimodal
MDF characteristic of most gE's would build up in two major starbursts:
the first burst forms the metal-poor clusters and the metal-poor
halo field stars, leaving most of the gas unused.  Perhaps one or
two Gyr later, a second major
burst uses up most of the gas and builds most of the metal-rich
stars along with the metal-richer clusters (\cite{for97};
\cite{har98}; \cite{glh99}).  The metal-poor component in this
picture should be more spatially extended, since it is visualized
to form while the protogalaxy was still in a clumpy, diffuse state.
Here again, we have
difficulty understanding the lack of metal-rich clusters, which we would
ordinarily have expected to form in the second burst.

\subsection{The Metal-Poor Clusters:  Normal or Abnormal?}

Any interpretation of this unusual galaxy must, at this point,
involve a healthy component of speculation.  In the spirit of
constructing a consistent picture, we suggest that {\it all} of the
principal anomalies in this galaxy (the narrow and entirely metal-poor
MDF; the rather low specific frequency; and the extremely extended
spatial distribution) are connected.  To put this statement another way,
let us pose the following question:  
what changes would be necessary to make the NGC 4874 GCS
resemble the ones that are conventionally regarded as ``normal'',
such as in the Virgo giants NGC 4472 and M87?  

For comparison purposes, let us look more closely at the data for
one of the most well studied giant ellipticals, NGC 4472, in which
the GCS has a distinctly bimodal MDF and an average $S_N \simeq 5$.
The excellent photometric study of \cite{lee98} shows
unequivocally that the metal-rich (red) and metal-poor (blue)
subsystems in NGC 4472 have different spatial distributions.  
A very similar phenomenon appears to hold in M87 (\cite{lee93};
\cite{kun99}), though with a larger total cluster
population (higher specific frequency).
It is therefore interesting to ask whether or not the {\it bluer}
cluster populations in these comparison galaxies have a characteristic
spatial extent (core radius) similar to what we find in NGC 4874.

\begin{figure}
\epsscale{0.8}
\plotone{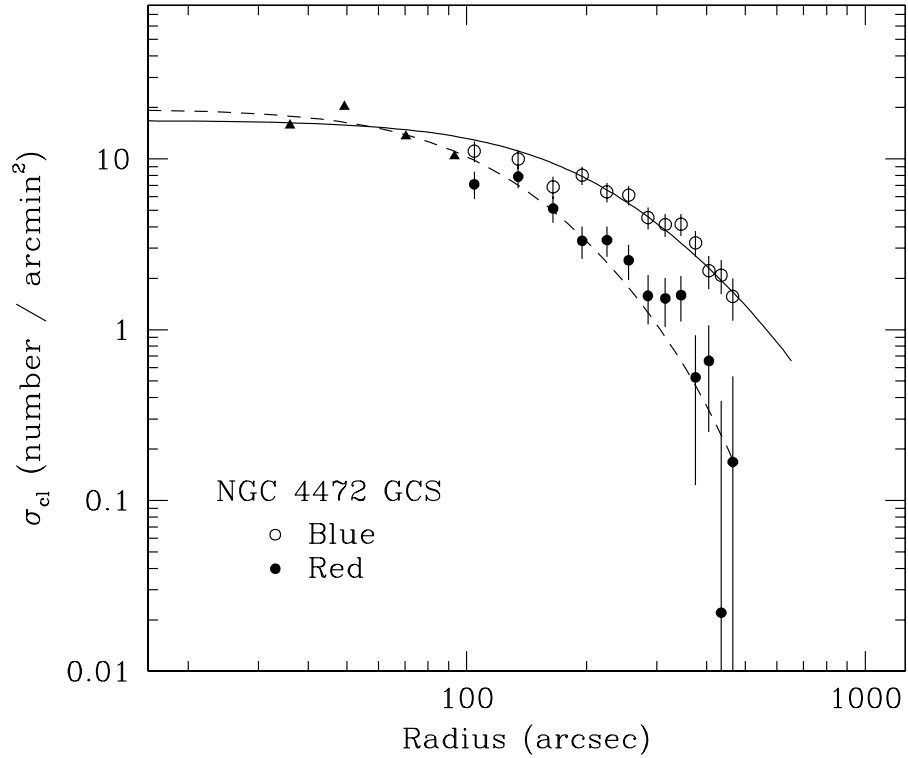}
\caption{Radial distribution for the globular cluster populations
in the Virgo elliptical NGC 4472.  Projected surface density (number
of clusters per arcmin$^2$) is plotted against mean radius
$r$(arcsec) for several annuli covering the range $r\simeq 2.3 -
40$ kpc.  {\it Solid symbols} are the metal-rich (redder) clusters,
open symbols the metal-poor (bluer) clusters.  Data are taken from
Lee \etal\ (1998), except for the small triangles at upper left,
which are from Harris \etal\ (1991).  The lines superimposed on each
component are King-model fits with parameters as given in the text.
}
\label{fig:NGC4472}
\end{figure}

\placefigure{fig:NGC4472}

The breakdown for NGC 4472 is shown in Figure \ref{fig:NGC4472}.
Here the projected density profiles are displayed separately for the
metal-poor (blue) and metal-rich (red) populations, along with
King-model fits to each.
The data for $r > 100''$ are from \cite{lee98},
including clusters brighter than $T_1 = 23$ and with
the divisions between the red and blue modes as prescribed in their
paper.  Data for the inner 
region $r < 100''$ are from \cite{hapv91},
multiplied by a factor 0.7 to correct for their different photometric
limit, and multiplied by a further factor of 0.5 with the assumption that
there are roughly equal numbers of blue and red clusters in 
this inner region.   
For the blue population, we find a core radius $r_c \simeq 184''$
(or 14 kpc for a distance modulus $(m-M)_0 = 31.0$) and central
concentration index $c \simeq 1.2$.  For the red population,
$r_c \simeq 100''$ (8 kpc) and $c \simeq 1.0$.  
The curve fits should be taken only as illustrative of the general 
shapes of the two subsystems, since the inner regions are
not well determined and the outermost background count level is also
somewhat uncertain (see Lee \etal\ 1998).  Nevertheless, 
its {\it low-metallicity}
cluster population has a radial extent not unlike the GCS 
in NGC 4874.

In summary, the {\it metal-poor} cluster population in 
NGC 4874 does not appear to be abnormal in total numbers, mean
metallicity, or spatial
structure by comparison what can be found in other giant ellipticals.
The missing metal-rich component is the key to understanding this system.
The most obvious single step to convert NGC 4874 into a more normal cD
would be simply {\it to add a roughly equal number of metal-rich 
clusters in the inner part of the galaxy}.  We would then have 
a cD galaxy with a 
bimodal MDF, a comfortably high specific frequency $S_N \sim 7$, 
and a GCS with subcomponents
that have distinct spatial distributions as we would normally expect.
The one uncomfortable anomaly that we would be left with, of course,
would be a higher-than-average mass ratio $\epsilon$.  But we do not
yet have a comprehensive set of measurements of $\epsilon$,
and it remains
to be seen whether or not this ratio is a truly universal one.

\subsection{Assembling the Metal-Poor Component}

An additional and possibly related piece of evidence is 
that the Coma cluster as a whole appears to have
assembled by the merger of smaller subgroups of galaxies.
Some of the subcomponents can still be detected
in the lumpiness of the X-ray contours on Megaparsec-size scales,
and in the redshift and spatial
distributions of the galaxies (e.g., \cite{mel88}; \cite{bur94};
\cite{whi93}; \cite{biv96}; \cite{sec96}; \cite{col96}; \cite{vik97};
\cite{con98}, among others).
The Coma cluster in total is so large, and its two central supergiant
galaxies so big, that these supergiants could well have experienced
a higher proportion of accretions or mergers than
in other giant ellipticals.  Such a process, in its entirety, would 
use a
mixture of all three classic galaxy formation scenarios ({\it in situ},
accretion, merger):  we could plausibly expect that
star formation would be taking place vigorously
in the dense gas of the central
proto-cD at the same time as large gaseous fragments and other
partially formed protogalaxies were raining in towards it.  

What constraints can be placed on a scenario of this type
from only the {\it metal-poor} cluster population?   It has already been
pointed out (see above) that the population of $\sim 9000$ total
clusters has an MDF
with the same mean metallicity and dispersion as that of a typical
halo GCS in a spiral or dwarf galaxy.  Possible interpretations for
their origin might then be the following:

\noindent
(a) They formed predominantly {\it in situ} during the 
first round of star and cluster formation in the protogalaxy.
In such a view, this epoch must have happened so early that
the protogalactic gas was still very clumpy and spatially
extended, and nearly unenriched.  Its traces are now seen in
the low-metallicity cluster population with its extremely large
core radius, but {\it not} in the main
bulk of the galaxy (which is dominated by a redder, much more
metal-rich and more centrally concentrated stellar population).
Accretion and stripping from small satellites might
have added more clusters and some metal-poor light later on,
building up the outer halo as it is seen today.

\noindent
(b) They formed predominantly in smaller, metal-poor
galaxies which were then accreted by NGC 4874.  
Many accretions must have been involved, 
combining a wide mixture of dwarfs and disk-type galaxies
more or less in the manner employed in the model by C\^ot\'e \etal\ (1998).

Let us explore this latter approach in a bit more detail.
The near-complete lack of clusters more metal-rich than about
[Fe/H] $\simeq -0.5$ ($V-I = 1.1$) suggests that not much accreted
material came from other fully formed {\it giant}
elliptical galaxies, since these would almost certainly 
have added metal-rich globulars to the GCS, which is contrary to
what we observe.  More quantitatively, we can ask 
how many clusters might have been accumulated from smaller galaxies
without leading to contradictions in either the GCS metallicities or
the (red) integrated color of the halo light.

Published analyses 
of the dE galaxies in Coma show that the fainter dE's are depleted
in the core relative to the brighter dE's and giant ellipticals
(\cite{lob97}; \cite{sec97}; \cite{tho93}; \cite{ber95};
\cite{tre98}).  The effect is most
strongly evident for dwarfs in the magnitude range $19 \ltsim R \ltsim 22.5$ 
(or $-15.4 \ltsim M_V \ltsim -11.9$, assuming $(V-R) = 0.6$ for a typical dE).
These intermediate-luminosity dE's have a nearly 
flat projected density distribution
in the Coma core ($r \ltsim 10'$), whereas 
the bright Coma ellipticals follow a steeper distribution
distribution $\sigma \sim r^{-1}$ closer
in to the center (\cite{ber95}; \cite{tre98}).  Interestingly, 
the faintest dE's ($R \gtsim 22.5$) appear to show
a centrally concentrated radial distribution more like the giants
(\cite{ber95}).  In addition, the luminosity
function of the dE's may be somewhat steeper in the outer regions of
Coma than in the core, suggesting again that depletion has occurred
in the core (\cite{lob97}; but see \cite{tre98}).

From the radial profile data of Secker \etal\ (1997) 
and Bernstein \etal\ (1995), we find that adding in about 850 
dE's in the faint range $19 < R < 23$
within $r < 8'$ (250 kpc) would be sufficient to bring their
radial distribution back up to the fiducial $\sigma(r) \sim r^{-1}$
characterized by the brighter ellipticals.  This number is about 1.5
times the {\it present-day} population of faint dE's in the same region.
The slightly brighter dE's ($R < 19$) also exhibit a flat
distribution $\sigma \sim r^{-0.5}$, though less so than does the
faintest group.  Increasing their numbers by $\sim 50 - 75$ (or 50
to 100 percent of their present-day core population) would bring
them back to the fiducial $r^{-1}$ profile as well.
In short, it seems plausible to suggest from the radial distribution
data that roughly 1000 small galaxies may have been removed from the
Coma core region.  

For purposes of building a strawman argument,
let us now assume that these ``missing'' galaxies were indeed
present at earlier times but were absorbed by
the central supergiant.  We can then assess how they would have affected
its globular cluster system and halo light.  
In the same manner as
C\^ot\'e \etal\ (1998), we perform a numerical experiment in which
these effects are built in:  

\begin{itemize}
\item
We assume, as above, that 1000
galaxies are accreted, and that they follow the composite 
luminosity distribution shape
given by Secker \& Harris (1996). The normal ellipticals follow 
a Gaussian number distribution peaked at $R=14.5$ with dispersion
$\sigma(R) = 0.80$, while the dwarf ellipticals follow a Schechter
function with $R^{\star}=15.6$ and power-law slope $\alpha = -1.4$.
\item
We assume that all the galaxies have a ``normal''
specific frequency $S_N=4$ (\cite{dur96}; \cite{mil98};
\cite{har98}), which fixes
their total contribution to the globular cluster population once the
luminosity is specified.
\item
We assume that the integrated color (i.e., metallicity) of each
galaxy increases with its luminosity according to the observed
correlation for Coma dwarfs (\cite{sec97}), $(B-R) = 2.41 - 0.056
R$.  The $(B-R)$ index can be converted to $(V-I)$ through
$(V-I)_0 \simeq 0.58 (B-R)_0 + 0.24$, valid for the Milky Way globular
clusters.
\item
Finally, we assume that the mean metallicity of the {\it globular
clusters} in each dwarf is correlated with galaxy luminosity
according to the relation derived by C\^ot\'e \etal\ (1998),
[Fe/H] $ = 2.31 + 0.638 M_V + 0.0247 (M_V)^2$.
Although bigger and brighter galaxies have more metal-rich clusters
on average, the mean [Fe/H] is only a weak function of $M_V$ for
the majority of the dwarfs (see also \cite{for97}).  In addition,
the average metallicity of the {\it clusters} is lower than that of the
{\it galaxy} they are in by typically $\sim 0.5$ dex (\cite{mil99}).
Thus, the combined galaxies can have a rather metal-poor GCS while
their aggregated halo light is somewhat more metal-rich.
\end{itemize}
The final parameter we employ is $M_V$(max), the upper cutoff
to the LF (that is, the luminosity of
the {\it brightest} accreted galaxy).  In summary, our numerical
procedure is to assume a value
for $M_V$(max), and add together 1000 galaxies fainter than this limit
which are distributed according to the rules listed above.  We then
ask how many globular clusters these would contribute, what their
mean metallicity will be, and what color the combined halo light
will have.  Larger accreted
galaxies are rarer, but have more populous GCSs and
redder (more metal-rich) clusters and halo light.  

\begin{figure}
\epsscale{0.8}
\plotone{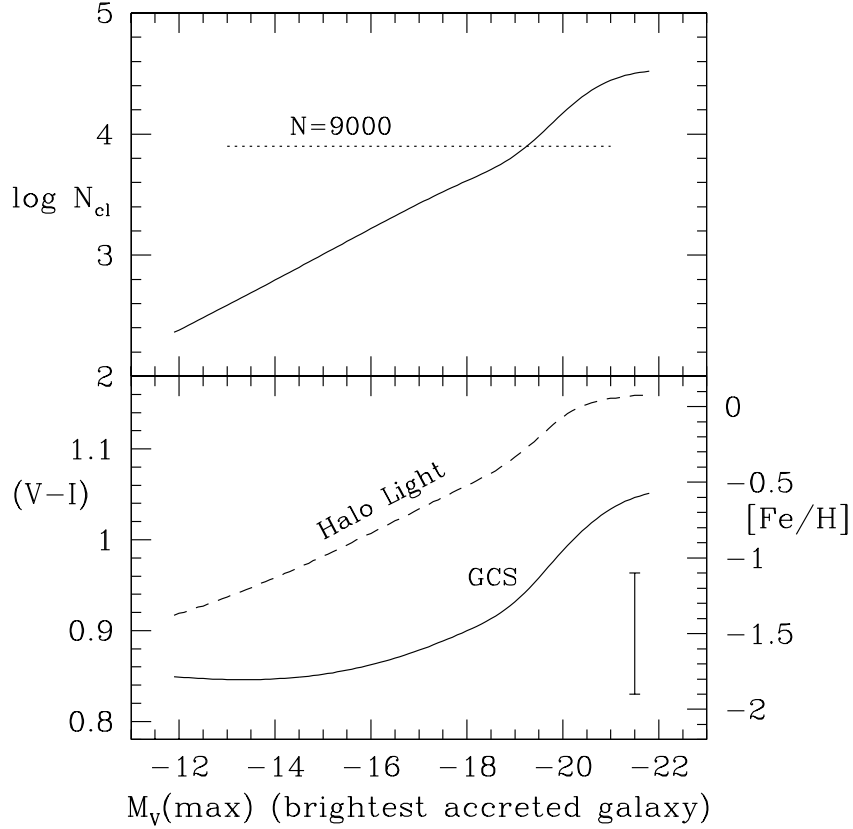}
\caption{Numerical model for an accreted globular cluster population. 
Here it is assumed that NGC 4874 has accreted
1000 galaxies fainter than $M_V$(max) which follow the Schechter-like
luminosity function given in the text.  The upper panel shows the
total number $N_{cl}$ of accreted globular clusters from these galaxies,
assuming that they all have a normal specific frequency $S_N = 4$.
In the lower panel, the solid line shows the mean [Fe/H] for the 
accreted clusters, with the observed metallicity range of the NGC 4874
clusters shown by the large error bar at right.  The dashed line shows
the mean $(V-I)$ color of the accreted {\it halo light}.
}
\label{fig:sumcl}
\end{figure}

\placefigure{fig:sumcl}

The results of this numerical exercise are summarized in Figure
\ref{fig:sumcl} as a function of the upper cutoff $M_V$(max).
(Rather similar information is contained in the more extensive
Monte Carlo models discussed
by \cite{cot98}; see especially their Figures 6 and 7).
The most important single trend to note is that all the output 
quantities (the number of accreted clusters $N_{cl}$, their mean 
[Fe/H], and the mean halo color $(V-I)$) are strongly sensitive to 
how many {\it bright} galaxies are in the sample.  Although the
luminous galaxies 
are individually rare, each one contributes so much total material that 
they tend to dominate the integrated sums.

These results can be used to place rough limits on the degree of
accretion  that NGC 4874 has experienced.
It is unlikely that many galaxies more luminous
than $M_V \sim -19$ have been accreted, since these would drive
the mean GCS metallicity to unacceptably high levels (the actual
MDF is indicated on the right side of the lower panel).  
At nearly this same $M_V$(max), we would also run into trouble with
the total number of accreted clusters, which would approach the
permitted maximum of $\sim 9000$.
However, if we stay within the allowed margin $M_V$(max) $\gtsim
-19$ to $-18$, the color of the halo light
contributed by the accreted objects stays within $(V-I) \simeq 1.0 \pm
0.1$.  It is an important consistency test that this color range
is comfortably within the observed range for the outer halo
of NGC 4874 (Fig.~\ref{fig:radcolor}).  

In summary, the accretion model yields a fairly
broad range of possibilities which would fit within the
observational constraints that we have.  With the assumptions
as listed above, we would suggest
that as much as half of the total
cluster population could have originated by accretion. 
Such a model would simultaneously be consistent with (a) the 
MDF of the clusters as we see it today, (b) the extended radial
distribution of the GCS, (c)  the color of the outer-halo
light, and (d) the present-day depletion of the 
dE's in the central Coma region.  

The first-order model we have just described does not,
by any means, lay out the full range of possibilities that could be
explored in more extensive simulations. (For example,
the Schechter LF exponent for the input dE's could be varied, 
as could their specific frequencies.  Both of these parameters would
be capable of generating noticeable differences around the mean
lines shown in Fig.~\ref{fig:sumcl}.)  It is precisely
this range of possibilities which makes it
difficult to answer just how much of the galaxy is accreted material,
as opposed to stars converted from {\it in situ} gas.
Because most
galaxies of all types -- dwarf ellipticals, irregulars, spirals,
and giant ellipticals -- contain a subpopulation of
metal-poor globular clusters 
([Fe/H] $\sim -1.6$ to $-1.0$) with only a weak dependence on
galaxy size, we cannot easily tell where such
clusters might have come from, except to rule out very luminous
parents.  

\subsection{The Strange Case of the Missing Red Clusters}

The foregoing arguments lead us to conclude that accretion of 
smaller satellites {\it may} have played an important
role in the history of this Coma supergiant.  But
the accretion (passive-merger) model cannot explain 
the presence of the much redder, more metal-rich material ($(V-I) \simeq 1.2$;
see Fig.~\ref{fig:radcolor}) in the central $\sim 20$
kpc of NGC 4874 which dominates its total light.  For this, we need
to resort either to {\it in situ} second-generation star formation
or to very gas-rich mergers which would be capable of driving the mean
metallicity up to its observed high levels.
Star clusters probably also formed during this stage as part of the
general conversion of gas into stars, since star clusters are 
found almost universally in star-forming regions (e.g., \cite{elm99}).  
Why are none of those clusters visible today?  To avoid appearing
in our photometric survey, any such objects
would have to be fainter than $V \sim 28$, or less massive than
about $10^5 M_{\odot}$ for a normal globular cluster age.

The formation of massive, globular-sized clusters ($10^5 - 10^6
M_{\odot}$) is likely to require local reservoirs of gas in the
range of $\sim 10^8 M_{\odot}$ or more (Searle-Zinn fragments or
supergiant molecular clouds; e.g., \cite{sea78}; \cite{lar93};
\cite{har94}).  By implication, giant molecular clouds (GMCs) 
this large may therefore not have been present in NGC 4874 during the stage
when most of the galaxy was being assembled.  
If, during this main star-forming stage, NGC 4874 was simultaneously
experiencing a wide range of tidal shocking, gas infall, merging,
supernova shocks, and internal winds, then the supergiant GMCs that
would normally make up the protogalaxy might have been extensively
broken down into smaller fragments.  With this line of
reasoning, we would have to postulate either that
these processes were more violent for NGC 4874 than in M87 or other giant
ellipticals with metal-rich clusters; or, that the timing of 
these events was such that they were particularly effective at
disrupting the large gaseous fragments.  Then, almost all
the star formation took place within these smaller GMCs. 

An evolutionary picture along these lines would imply that metal-rich star
clusters did form along with the main bulk of the galaxy, but that
they were in a mass range $\ltsim 10^5 M_{\odot}$ which would
fall below the limits of our photometric survey.
We can further speculate that these smaller star clusters -- more
easily subject to dynamical destruction than the massive ones --
have to a large extent already dissolved into the general field,
leaving few traces in the GCS that we see today.

An additional observation which is, perhaps, consistent with a such
a view is that the active mergers between disk galaxies which
we can directly observe today (e.g., \cite{zep99}; \cite{ws95}; 
\cite{whi99})
seem to be efficient at producing large numbers of {\it low-mass}
star clusters.  (For example, the well known Antennae merger has
generated well over 1000 low-mass ($M \ltsim 10^5 M_{\odot}$)
clusters but only a few dozen massive young globulars.)
For this reason, these same mergers will generate
ellipticals with rather modest specific frequencies in the range
$S_N \sim 2 - 3$ (\cite{har99} and the papers
listed above), since it is the high-mass
clusters which survive longest and determine the long-term value of $S_N$.

It is obvious that the storyline we have just proposed has
strong elements of ``special pleading''; that is, arguments
constructed specifically for one situation.  But NGC 4874 
presents us with a unique and outstanding set of GCS
features within a rare type of galaxy, and we may in the end
be forced to deal with it in a different way than with 
more normal ellipticals.

\section{Summary}

HST/WFPC2 photometry in $V$ and $I$ has been obtained for the
globular cluster system in NGC 4874, the central cD in Coma.
The luminosity function (GCLF) of the clusters has the normal
Gaussian-like shape and turnover level, but several other features
of the system prove to be surprising.  The GCS is spatially very
extended, with core radius $r_c \sim 22$ kpc, and the metallicity
distribution function is unimodal, narrow ($\sigma$[Fe/H] $\simeq
0.4$), and entirely metal-poor ($\langle$[Fe/H]$\rangle \sim -1.5$).
Lastly, the total population $N_t \simeq 9000 \pm 1000$ gives a
specific frequency $S_N = 3.7 \pm 0.5$ which is in the typical range
for normal ellipticals but strikingly low for central cD-type galaxies.

We suggest that the principal anomaly in this GCS is essentially the
complete lack of metal-rich clusters.  If these were present in
normal (M87-like) numbers in addition to the metal-poor ones
that are already there, then the GCS in total would closely
resemble what we see in many other dominant cD galaxies.
This supergiant galaxy, in its early stages, appears to have avoided forming
globular clusters during the main metal-rich stage of star formation
which built the bulk of the galaxy.  This situation presents strong
challenges to all three of the classic modes of galaxy formation:
accretion, disk mergers, and {\it in situ} formation.
As a best-compromise interpretation, we suggest that up to half the
cluster population could have been gained by the accretion of small
satellite galaxies which richly populated the Coma core region
at earlier times.
But the main, metal-rich star formation stage which built
the inner parts of the galaxy must have somehow avoided generating
massive star clusters.   We suggest that massive high-metallicity
globulars did not form because the 
normal sites of globular cluster formation -- supergiant molecular
clouds -- had
already been broken down into smaller clouds at the time the
metal-rich star formation was taking place.

\acknowledgements

This research was supported through grants from
the Natural Sciences and Engineering Research Council of Canada.
We are grateful to Dean McLaughlin for supplying the deprojection
code.

{}

\begin{deluxetable}{rlrcr}
\tablewidth{0pt}
\tablecaption{\label{tab:rad}GCS Radial Profile Data }
\tablehead{
\colhead{$r$} & \colhead{n} & \colhead{Area } & \colhead{$\sigma$} &
\colhead{$\rho$} \\
\colhead{(arcsec)} & & \colhead{(arcsec$^2$)} & \colhead{(arcsec$^{-2}$)}
& \colhead{($M_{\odot}$ kpc$^{-3}$)} \\
}
\startdata
    9.9  &~$86.86 \pm 9.94$& 236.06 &    $0.3680 \pm   0.0421$ & 25800\\
    13.6  &~$82.46 \pm 9.70$& 236.55 &    $0.3486 \pm   0.0410$ & 39400 \\
    17.9  &~$60.06 \pm 8.31$& 212.65 &    $0.2824 \pm   0.0391$ & 19700 \\
    19.3  &~$94.64 \pm10.44$& 368.27 &    $0.2570 \pm   0.0283$ & 12700 \\
    27.6  &$133.16 \pm12.34$& 558.76 &    $0.2383 \pm   0.0221$ & 12400 \\
    35.5  &$163.34 \pm13.64$& 798.82 &    $0.2045 \pm   0.0171$ & 9350 \\
    43.4  &$169.79 \pm13.93$& 981.35 &    $0.1730 \pm   0.0142$ & 7090 \\
    51.2  &$187.49 \pm14.57$&1254.53 &    $0.1494 \pm   0.0116$ & 3040 \\
    55.2  &$402.11 \pm21.39$&2811.67 &    $0.1430 \pm   0.0076$ & 4540 \\
    70.3  &$388.91 \pm21.04$&3396.13 &    $0.1145 \pm   0.0062$ & 3140 \\
    85.7  &$298.71 \pm18.47$&3263.40 &    $0.0915 \pm   0.0057$ & 3070 \\
   100.0  &$107.63 \pm11.08$&1739.88 &    $0.0619 \pm   0.0064$ & 1470 \\
   119.1  &~~$25.95 \pm 5.40$& 496.25 &   $0.0523 \pm   0.0109$ & 520 \\
\enddata
\label{tab:radprof}
\end{deluxetable}

\begin{deluxetable}{rrrr}
\tablewidth{0pt}
\tablecaption{Surface Photometry for NGC 4874}
\tablehead{\colhead{$r$ ($''$)} & \colhead{$(V-I)$} &
\colhead{$r$ ($''$)} & \colhead{$(V-I)$} \\
}
\startdata 
         1.33    &   1.289 & 9.65    &   1.225 \\
         2.79    &   1.290 & 16.52   &   1.213 \\
         2.92    &   1.303 & 17.97   &   1.192 \\
         2.99    &   1.298 & 19.25   &   1.215 \\
         3.23    &   1.289 & 22.00   &   1.180 \\
         3.43    &   1.287 & 27.00   &   1.180 \\
         3.96    &   1.268 & 32.00   &   1.155 \\
         4.18    &   1.265 & 35.00   &   1.130 \\ 
         4.53    &   1.285 & 60.00   &   1.080 \\
         5.24    &   1.247 & 95.00   &   0.930 \\ 
         7.51    &   1.227 \\
\enddata
\label{tab:surface}
\end{deluxetable}

\begin{deluxetable}{lcccc}
\tablewidth{0pt}
\tablecaption{Aperture Corrections for Photometric Calibration}
\tablehead{\colhead{Source} & \colhead{$\Delta V$ (WF)} & 
\colhead{$\Delta I$ (WF)} & \colhead{$\Delta V$ (PC1)} & \colhead{$\Delta I$ (PC1)} \\
}
\startdata 
 This study & 0.25 & 0.31 & 0.24 & 0.31 \\
 Holtzman \etal\ 1995a & 0.19 & 0.23 & 0.20 & 0.23 \\
 Suchkov \& Casertano 1997 & 0.23 & 0.21 \\
\enddata
\tablenotetext{}{
{\it NB:} For the WF chips, $\Delta m$ is the magnitude difference
between the 2-px and 5-px apertures.  For the PC1 chip, it is the
difference between the 3-px and 11-px apertures.}
\label{tab:apcorr}
\end{deluxetable}

\clearpage

\clearpage

\end{document}